\documentclass[letterpaper,english,aps,showkeys]{revtex4-1}
\usepackage[T1]{fontenc}
\usepackage[latin9]{inputenc}
\usepackage{color}
\usepackage{amsmath}
\usepackage{amssymb}



\usepackage{babel}
\begin{document}

\title{Exact solutions of scalar bosons in the presence of the Aharonov-Bohm
and Coulomb potentials in the gravitational field of topological defects}

\author{Abdelmalek Boumali}

\affiliation{Laboratoire de Physique Appliquée et Théorique, ~\\
University Larbi Tébessi -Tébessa-, 12000, W. Tébessa, Algeria.}
\email{boumali.abdelmalek@gmail.com}

\author{Houcine Aounallah}

\affiliation{Laboratoire de Physique Appliquée et Théorique, ~\\
University Larbi Tébessi -Tébessa-, 12000, W. Tébessa, Algeria.}
\email{Houcine12400@gmail.com}

\begin{abstract}
In this paper, we analyze the relativistic quantum motion of a charged
scalar particles in the presence of a Aharonov\textendash Bohm and
Coulomb potentials in the spacetimes produced by an idealized cosmic
string and global monopole. We have calculated and discussed the eigensolutions
of DKP equation and their dependence in both the geometry of the spacetimes
and coupling constants parameters.
\end{abstract}

\keywords{DKP equation; Aharonov-Bohm potential; Topological defects}
\maketitle

\section{INTRODUCTION}

The analysis of gravitational interactions with a quantum mechanical
system has recently attracted attention in particle physics and has
been an active field of research.The general way to understand the
interaction between relativistic quantum mechanical particles and
gravity is to solve the general relativistic form of their wave equations.
These solutions are valuable tools for examining and improving models
ans numerical methods for solving complicated physical problems.

The DKP equation is a relativistic wave equation which describes spin-0
and spin-1 particles, and it is a direct generalization to the Dirac
particles of integer spin in which one replaces the gamma matrices
by beta metrics, where the four beta matrices satisfy the Kemmer algebra
\citep{1,2,3,4}. \textcolor{red}{One important question related to
the DKP equation concerns the equivalence between its spin 0 and 1
sectors and the theories based on the second-order KG and Proca equations.
The Dirac-like DKP equation is not new and dates back to the 1930s.
Historically, the loss of interest in the DKP equation stems from
the equivalence of the DKP approach to the KG and Proca descriptions
in on-shell situations, in addition to the greater algebraic complexity
of the DKP formulation. However, in the 1970s, this supposed equivalence
was questioned in several situations involving breaking of symmetries
and the hadronic possess, showing that in some cases, the DKP and
KG theories can give different results. Moreover, the DKP equation
appears to be richer than the KG equation if interactions are introduced. }

The topological defects plays an important role in physical properties
of systems, and they appear in gravitation as monopoles, strings and
walls. Among them, cosmic string and monopole seem to be the best
candidates to be observed. Cosmic strings and global monopoles are
exotic topological objects: they do not produce local gravitational
interaction; however, they modify the geometry of the spacetime, producing
planar and solid angle deficit respectively\citep{5}.

The effect of the Aharonov-Bohm potential (\textbf{AB}) is a purely
quantum phenomenon which has no classical equivalent. The eminent
work of \textbf{AB} in 1959 revealed that the electromagnetic potentials
are in reality fundamental physical objects that can manifest themselves
in matter where the corresponding fields exert no force (Lorentz's
force is zero )\citep{6}. They proposed an experiment to test, in
quantum mechanics, the coupling of electric charges to electromagnetic
field strengths through a local interaction with the electromagnetic
potential \textbf{A}, but not with the field strengths themselves.
Also, they predicted a quantum interference effect due to the motion
of charged particles in regions where \textbf{B}(\textbf{E}) vanishes,
but not \textbf{A}, leading to a nonlocal gauge-invariant effect depending
on the flux of the magnetic field in the inaccessible region, in the
magnetic case, and on the difference of the integrals over time of
time-varying potentials, in the electric case \citep{7}.

The eigenproblem of the Kinetic Angular Momentum (KAM) of the electron
in the Aharonov\textendash Bohm effect has been solved by Kretzschmar
\citep{8}. The total Hilbert space of the eigenfunctions is split
into two subspaces and the symmetry of the motion of the electron
around the magnetic flux makes Pauli's criterion inapplicable \citep{9,10,11,12}:
that means the condition of admissibility of the wave function in
the region of AB potential. In this way, Henneberger \citep{9}, based
on the Pauli criterion conclude that (i) the single valuedness imposed
by Aharonov and Bohm is not the correct one, and consequently, (ii)
the \textbf{AB} effect, if it exists, may not be interpreted as scattering.
Roy and Singh \citep{10}and Chun-Fang Li \citep{11,12} showed that
the Pauli criterion cannot be applied in the time-dependent: we note
here that the inapplicability of the Pauli criterion means that -(i)
the KAM does not satisfy the fundamental commutation relations of
the angular momentum, and (ii) it reflects the breakdown of the symmetry
of the electron's motion around the solenoid. Thus, as argued by Chun-Fang
Li \citep{11,12}, in the presence of \textbf{AB} potential, the KAM
of the electron does not satisfy the fundamental commutation relations
of the angular momentum even when the electron is restricted to the
doubly-connected space where it does not touch the magnetic field
on the z-axis: the region where the magnetic field exists and is inaccessible
to the electron should also be taken into account in the physically
meaningful commutation relations. This situation has been used in
the case of the interaction of the spin 0 particle with an \textbf{AB}
potential of two and three dimensions by the application of the Duffin\textendash{}
Kemmer\textendash Petiau equation in Minkowski space-time \citep{13}.

The principal aim of this paper is to solve the DKP equation in the
presence of a Coulomb and \textbf{AB} potentials in the gravitational
field of topological defects. In the presence of \textbf{AB} potential,
we will show that that the for admissibility of the wave function
is inapplicable \citep{8,9,10,11,12}. The inapplicability of the
Pauli criterion reflects the breakdown of the symmetry of the particle's
motion around the solenoid \citep{11,12}.

\textcolor{red}{The outline of this paper is as follows: in Sec. II,
a generalized DKP equation is introduced in a curved space-time. In
section 3, the eigenfunctions of scalar bosons in the presence of
the Aharonov-Bohm and Coulomb potentials in the gravitational field
of topological defects such as cosmic string and global monopole have
been discussed in Sec. III and IV. Finally, Sec. V will be a conclusion.}

\section{The DKP equation in curved space}

\subsection{The formalism}

The first-order relativistic DKP equation for a free charged scalar
bosons of mass $M$ in flat space-time is given by \citep{1,2,3,4}
\begin{equation}
(i\beta^{\mu}\partial_{\mu}-\frac{Mc}{\hbar})\psi=0,\label{eq:1}
\end{equation}
where $\beta^{\mu}(\mu=0,1,2,3)$ are the DKP matrices which satisfy
the following commutation rules
\begin{equation}
\beta^{\kappa}\beta^{\nu}\beta^{\lambda}+\beta^{\lambda}\beta^{\nu}\beta^{\kappa}=g^{\kappa\nu}\beta^{\lambda}+g^{\nu\lambda}\beta^{\kappa},\label{eq:2}
\end{equation}
and $g^{\mu\nu}=\text{diag}\left(1,-1,-1,-1\right)$ is the Minkowski
metric tensor. For the flat space, the beta matrices are chosen as
follows \citep{14} 
\begin{equation}
\beta^{0}=\left(\begin{array}{cc}
\nu & \tilde{0}\\
\tilde{0}_{T} & \mathbf{0}
\end{array}\right),\beta^{i}=\left(\begin{array}{cc}
\hat{0} & \rho^{i}\\
-\rho_{T}^{i} & \mathbf{0}
\end{array}\right),\label{eq:3}
\end{equation}
with $\hat{0}$, $\tilde{0}$, $\mathbf{0}$ as $2\times2$, $2\times3$,
$3\times3$ zero matrices, respectively and
\begin{equation}
\nu=\left(\begin{array}{cc}
0 & 1\\
1 & 0
\end{array}\right),\rho^{1}=\left(\begin{array}{ccc}
-1 & 0 & 0\\
0 & 0 & 0
\end{array}\right),\rho^{2}=\left(\begin{array}{ccc}
0 & -1 & 0\\
0 & 0 & 0
\end{array}\right),\rho^{3}=\left(\begin{array}{ccc}
0 & 0 & -1\\
0 & 0 & 0
\end{array}\right).\label{eq:4}
\end{equation}
In curved space-time, Eq. (\ref{eq:1}) is written by \citep{15,16,17}
\begin{equation}
\left\{ i\tilde{\beta}^{\mu}\left(\partial_{\mu}+\frac{1}{2}\omega_{\mu ab}S^{ab}-\frac{ie}{\hbar c}A_{\mu}\right)-\frac{Mc}{\hbar}\right\} \Psi=0,\label{eq:5}
\end{equation}
\textcolor{red}{where $A_{\mu}$ denotes the vector potential associated
with the electromagnetic field,} $S^{ab}=\left[\beta^{a},\beta^{b}\right]$
and $\tilde{\beta}^{\mu}$ are the Kemmer matrices in curved space-time.
These matrices are related to their Minkowski counterparts, $\beta$,
via 
\begin{equation}
\tilde{\beta}^{\mu}=e_{(a)}^{\mu}\beta^{a}.\label{eq:6}
\end{equation}
The following tetrad relations and the spin connection are calculate
by using by the relation
\begin{equation}
\omega_{\mu ab}=e_{'(a)l}e_{(b)}^{j}\Gamma_{j\mu}^{l}-e_{(b)}^{j}\partial_{\mu}e_{'(a)j},\label{eq:7}
\end{equation}
where 
\begin{equation}
\Gamma_{\nu\lambda}^{\mu}=\frac{g^{\mu\rho}}{2}\left(g_{\rho\nu,\lambda}+g_{\rho\lambda,\nu}-g_{\nu\lambda,\rho}\right),\label{eq:8}
\end{equation}
are the Christoffel symbols \citep{18}.

Now, with the aids of equations (\ref{eq:6}), (\ref{eq:7}) and (\ref{eq:8}),
spin connection coefficients are :
\begin{itemize}
\item for the metric corresponding to a cosmic string with \citep{19,20,34,35}
\begin{equation}
ds^{2}=dt^{2}-dr^{2}-r^{2}d\theta^{2}-a'^{2}r^{2}sin^{2}\theta d\varphi^{2}\label{eq:9}
\end{equation}
where$-\infty<\text{t}<+\infty,$ $0\leqslant r$ , $0\leqslant\theta\leqslant\pi$,
and $0\leqslant\varphi\leqslant2\pi$, the $\omega_{\mu ab}$ are
\begin{equation}
\omega_{\theta ab}=\left(\begin{array}{cccc}
0 & 0 & 0 & 0\\
0 & 0 & 1 & 0\\
0 & -1 & 0 & 0\\
0 & 0 & 0 & 0
\end{array}\right),\omega_{\varphi ab}=\left(\begin{array}{cccc}
0 & 0 & 0 & 0\\
0 & 0 & 0 & a'\text{sin}\theta\\
0 & 0 & 0 & a'\text{cos}\theta\\
-a'\text{sin}\theta & -a'\text{cos}\theta & 0 & 0
\end{array}\right).\label{eq:10}
\end{equation}
Here the tetrad $e_{(a)}^{\mu}$ are chosen to be
\begin{equation}
e_{(a)}^{\mu}=\left(\begin{array}{cccc}
1 & 0 & 0 & 0\\
0 & 1 & 0 & 0\\
0 & 0 & \frac{1}{r} & 0\\
0 & 0 & 0 & \frac{1}{a'r\text{sin}\theta}
\end{array}\right).\label{eq:11}
\end{equation}
\textcolor{red}{The parameter $a'$ is the deficit angle associated
with conical geometry obeying $a'=1-4\eta$ , and $\eta$ is the linear
mass density of the string. It is defined in the range (0, 1{]} and
corresponds to a deficit angle $\varOmega=2\pi(1-\alpha)$. From the
geometrical point of view, the metric in Eq. (\ref{eq:9}) describes
a Minkowski space-time with a conical singularity \citep{38}.}
\item in the case of the global monopole with \citep{21,22,36,37}
\begin{equation}
ds^{2}=dt^{2}-dr^{2}-b'^{2}r^{2}\left(d\theta^{2}+\text{si\ensuremath{n^{2}}}\theta d\varphi^{2}\right),\label{eq:12}
\end{equation}
 we have
\begin{equation}
\omega_{\theta ab}=\left(\begin{array}{cccc}
0 & 0 & 0 & 0\\
0 & 0 & b' & 0\\
0 & -b' & 0 & 0\\
0 & 0 & 0 & 0
\end{array}\right),\omega_{\varphi ab}=\left(\begin{array}{cccc}
0 & 0 & 0 & 0\\
0 & 0 & 0 & b'\text{sin}\theta\\
0 & 0 & 0 & \text{cos}\theta\\
-b'\text{sin}\theta & -\text{cos}\theta & 0 & 0
\end{array}\right),\label{eq:13}
\end{equation}
where the tetrad $e_{(a)}^{\mu}$ are
\begin{equation}
e_{(a)}^{\mu}=\left(\begin{array}{cccc}
1 & 0 & 0 & 0\\
0 & 1 & 0 & 0\\
0 & 0 & \frac{1}{b'r} & 0\\
0 & 0 & 0 & \frac{1}{b'r\text{sin}\theta}
\end{array}\right).\label{eq:14}
\end{equation}
\textcolor{red}{The parameter $b'$, smaller than unity, depends on
the energy scale where the monopole is formed and where the symmetry
is broken. The spacetime described by (\ref{eq:12}) has a non-vanishing
scalar curvature, $R=\frac{2\left(1-b'\right)}{b'^{2}r^{2}}$ , and
presents a solid angle deficit $\Omega'=4\pi^{2}(1-b'^{2})$. The
many interesting investigations of physical effects associated with
global monopole consider this object as a point-like defect (see Refs
\citep{39,40} and references therein).}
\end{itemize}
Let's now discuss the problem of Pauli criterion in the presence of
\textbf{AB} potential for both cosmic string and global monopole cases.

\subsection{Pauli Criterion and the Aharonov-Bohm Effect: review}

As we know, the KAM satisfy the following relations 
\begin{equation}
\mathbf{L}\times\mathbf{L}=i\hbar\mathbf{L},\label{eq:15}
\end{equation}
with the following standard commutations rules \citep{23}
\begin{equation}
\left[L_{i},L_{j}\right]=i\hbar\epsilon_{ijk}L_{k},\label{eq:16}
\end{equation}
\begin{equation}
\left[L_{z},L_{\pm}\right]=\pm\hbar L_{\pm},\label{eq:17}
\end{equation}
\begin{equation}
\left[L^{2},L_{\pm}\right]=0,\label{eq:18}
\end{equation}
where
\begin{equation}
L^{2}=-\hbar^{2}\left\{ \frac{1}{\text{sin}\theta}\frac{\partial}{\partial\theta}\left(\text{sin}\theta\frac{\partial}{\partial\theta}\right)+\frac{1}{\text{sin}^{2}\theta}\frac{\partial}{\partial\varphi}\right\} ,\label{eq:18.1}
\end{equation}
and where
\begin{equation}
L_{z}=-i\hbar\frac{\partial}{\partial\varphi},
\end{equation}
with $L_{\pm}=L_{x}\pm iL_{y}$ are the ladder operators. The KAM
is based on these commutation relations that Pauli required that the
appropriate eigenfunction be those which are square integrable and
are closed under the operation of the ladder operators. This condition
is called as the Pauli criterion, and consequently the commutations
relations in Eq. (\ref{eq:15}) are pertinent for the Pauli criterion
to be applicable.

In the presence of the\textbf{ AB} potential where
\begin{equation}
A_{r}=A_{\theta}=0,\,\,\,\,\,A_{\varphi}=\frac{\Phi}{2\pi r\text{sin}\theta},
\end{equation}
the KAM becomes
\begin{equation}
L^{2}=-\hbar^{2}\left\{ \frac{1}{\text{sin}\theta}\frac{\partial}{\partial\theta}\left(\text{sin}\theta\frac{\partial}{\partial\theta}\right)+\frac{1}{\text{sin}^{2}\theta}\left(\frac{\partial}{\partial\varphi}-i\alpha\right)^{2}\right\} ,\label{eq:18.1-1}
\end{equation}
with
\begin{equation}
L_{z}=-i\hbar\left(\frac{\partial}{\partial\varphi}-i\alpha\right).
\end{equation}
\textcolor{red}{Here the particle is placed in the vector potential
field of an infinite long solenoid of magnetic flux $\Phi=\alpha\Phi_{0}$
where $\Phi_{0}=\frac{hc}{e}$ and the parameter $\alpha$ is the
amount of the magnetic flux.}

Chun-Fang Li \citep{11,12} shows that, in the presence of \textbf{AB}
potential, the KAM does not satisfy the fundamental commutation relations
(\ref{eq:15}), and instead of (\ref{eq:15}), he obtains
\begin{equation}
\mathbf{L}\times\mathbf{L}=i\hbar\mathbf{L+\left(\mathrm{2im}\alpha\hbar^{2}\right)\mathrm{cos\theta\mathbf{\delta\mathrm{\left(\text{cos}^{2}\theta-1\right)}}}\textrm{\ensuremath{\boldsymbol{e}_{r}}}}.\label{eq:19}
\end{equation}
From (\ref{eq:19}), both Eqs. (\ref{eq:17}) and (\ref{eq:18}) are
transformed into 
\begin{equation}
\left[L_{z},L_{\pm}\right]=\pm\hbar L_{\pm}\label{eq:17-1}
\end{equation}
\begin{equation}
\left[L^{2},L_{\pm}\right]=\mp2\alpha\hbar^{2}\left[L_{\pm}\delta\left(\text{cos}^{2}\theta-1\right)+\delta\left(\text{cos}^{2}\theta-1\right)L_{\pm}\right].
\end{equation}
These rules are different from Eqs. (\ref{eq:15}) and (\ref{eq:18}).
As described by Chun-Fang Li, the region where the magnetic field
exists is inaccessible to the electron, the commutation relations
of the KAM should take it into account. This type of commutations
relations as said to be global following Chun-Fang Li, and consequently
the Pauli criterion is innaplicable.

The eigensolutions of $L_{z}$ and $L^{2}$ under the following boundary
condition at $\theta=0,\pi$,
\begin{equation}
\psi\left(\theta,\varphi\right)_{\mid_{\theta=o,\pi}}=0.
\end{equation}
are presented by Kretzschmar\citep{8}. This condition means that
the particle is restricted to the doubly-connected region of $\theta\neq0,\pi$.
According to Kretzschmar, and with the following substitutions
\begin{equation}
L_{z}\psi_{l\lambda}\left(\theta,\varphi\right)=\lambda\hbar\psi_{l\lambda}\left(\theta,\varphi\right),
\end{equation}
\begin{equation}
L^{2}\psi_{l\lambda}\left(\theta,\varphi\right)=l\left(l+1\right)\hbar^{2}\psi_{l\lambda}\left(\theta,\varphi\right),
\end{equation}
\begin{equation}
\lambda=m-\alpha,\,\,\,\,\,m=\left(0,\pm1,\pm2,\ldots\right),
\end{equation}
\begin{equation}
l=\left|\lambda\right|+\kappa,\,\,\,\kappa=0,1,2,\ldots,
\end{equation}
the normalized eigenfunctions are of the form
\begin{equation}
\psi_{\lambda l}\left(\theta,\varphi\right)=c_{\lambda,l}P_{l}^{-\left|\lambda\right|}\left(\text{cos}\theta\right)e^{im\varphi},\,\,\,m=0,\pm1,\pm2,\ldots,
\end{equation}
where
\begin{equation}
c_{\lambda,l}=e^{\frac{i\pi}{2}\lambda+\frac{i\pi}{2}\left|\lambda\right|}\left(\frac{2l+1}{4\pi}\frac{\Gamma\left(l+\left|\lambda\right|+1\right)}{\Gamma\left(l-\left|\lambda\right|+1\right)}\right)^{\frac{1}{2}}.
\end{equation}
According to the sign of $\lambda$, we have
\begin{equation}
\begin{cases}
\psi_{\lambda_{1}l_{1}}\left(\theta,\varphi\right)=c_{\lambda_{1},l_{1}}P_{l_{1}}^{-\lambda_{1}}\left(\text{cos}\theta\right)e^{im\varphi}, & \lambda_{1}=m-\alpha>0,\,\,l_{1}=\lambda_{1}+\kappa,\\
\psi_{\lambda_{2}l_{2}}\left(\theta,\varphi\right)=c_{\lambda_{2},l_{2}}P_{l_{2}}^{\lambda_{2}}\left(\text{cos}\theta\right)e^{im\varphi}, & \lambda_{2}=m-\alpha<0,\,\,l_{2}=-\lambda_{2}+\kappa.
\end{cases}
\end{equation}
The parameters $\lambda$ and $l$ dependent on the amount $\alpha$
of the magnetic field which is not correct when we adopt the KAM as
written by Eq. (1). In addition the innaplicability of the Pauli criterion
modified completely the total Hilbert space S: Following Kretzschmar
\citep{8}, the total Hilbert space S is split into two subspaces,
$S_{+}$ and $S_{-}$. $S_{+}$ is spanned by all the eigenfunctions
$\psi_{\lambda_{1}l_{1}}\left(\theta,\varphi\right)$, and $S_{-}$
is spanned by all the eigenfunctions $\psi_{\lambda_{2}l_{2}}\left(\theta,\varphi\right)$.
These two subspaces are not connected by the ladder operators.

\textcolor{red}{In this stage, we are able to extend the works of
Kretzschmar \citep{8} (splitting of Hilbert space) and Chun-Fang
Li \citep{11,12} (the non applicability of Pauli's criterion) to
the case of scalar bosons in the gravitational field of topological
defects such as cosmic string and general monopole: first, we are
interested in deducing the $\mathbf{L}\times\mathbf{L}$ in the presence
of of topological defects, then solving the scalar DKP equation by
taking into account the works of both authors.}
\begin{itemize}
\item \textbf{\emph{case of cosmic string}}
\end{itemize}
In cosmic string, replacing the vector \textbf{AB} by the relation
\begin{equation}
A_{\varphi}=\frac{\Phi}{2\pi a'r\text{sin}\theta},\label{eq:20}
\end{equation}
Eq. (\ref{eq:15}) is transformed into
\begin{equation}
\mathbf{L}\times\mathbf{L}=i\hbar\boldsymbol{L}+\frac{2i\hbar^{2}\alpha}{a'}\mathrm{\text{cos}\theta\mathbf{\delta\mathrm{\left(cos^{2}\theta-1\right)}}}\boldsymbol{e}_{r}.\label{eq:21}
\end{equation}
 
\begin{itemize}
\item \textbf{\emph{case of global monopole}}
\end{itemize}
In the global monopole, with
\begin{equation}
A_{\varphi}=\frac{\Phi}{2\pi b'r\text{sin}\theta},\label{eq:22}
\end{equation}
Eq. (\ref{eq:15}) becomes 
\begin{equation}
\mathbf{L}\times\mathbf{L}=i\hbar\mathbf{L+\frac{\mathrm{2i\hbar^{2}\alpha}}{b'^{2}}\mathrm{cos\theta\mathbf{\delta\mathrm{\left(cos^{2}\theta-1\right)}}}\textrm{\ensuremath{\boldsymbol{e}_{r}}}}\label{eq:23}
\end{equation}
Both Eqs. (\ref{eq:21}) and (\ref{eq:23}) are obviously different
from Eq. (\ref{eq:15}). According to Chun-Fang Li \citep{11,12},
we note that in the presence of \textbf{AB} potential, the KAM of
the particle, in both cases, does not satisfy the fundamental commutation
relations of the angular momentum even when the particle is restricted
to the doubly-connected space where it does not touch the magnetic
field on the z-axis. The region where the magnetic field exists and
is inaccessible to the particle should also be taken into account
in the physically meaningful commutation relations.

\textcolor{red}{In what follow, we are ready to present the solutions
of scalar bosons by taking into account that the fundamental commutation
relations of the angular momentum are modified in both cosmic string
and general monopole cases.}

\section{The dkp equation with the aharonov-bohm and coulomb potentials in
the presence of a cosmic string.}

\subsection{The solutions}

The cosmic string space-time with an internal magnetic field in spherical
coordinates is described by the line element
\begin{equation}
ds^{2}=dt^{2}-dr^{2}-r^{2}d\theta^{2}-a'^{2}r^{2}\text{sin}^{2}\theta d\varphi^{2}.\label{eq:23.1}
\end{equation}
The 4-vector potential,in the background of a cosmic string, reads\textcolor{red}{
\begin{equation}
A_{0}=\frac{kq}{r},\begin{array}{cc}
A_{r}=0,\,A_{\theta}=0, & A_{\varphi}=\frac{\Phi}{2\pi a'rsin\theta}.\end{array}\label{eq:24}
\end{equation}
}From Eq. (\ref{eq:5}), and by using Eqs. (\ref{eq:3}), (\ref{eq:6})
and (\ref{eq:10}), the DKP equation in the presence of both \textbf{AB}
and Coulomb potentials is given by
\begin{equation}
\left[\frac{\beta^{0}}{\hbar c}\left(E-\frac{kq}{r}\right)+i\beta^{1}\partial_{r}+\frac{i\beta^{2}}{r}\left(\partial_{\theta}-\beta^{2}\beta^{1}\right)+\frac{i\beta^{3}}{a'r\text{sin}\theta}\left\{ \left(\partial_{\varphi}-i\alpha\right)-a'\text{sin}\theta\beta^{3}\beta^{1}-a'\text{cos}\theta\beta^{3}\beta^{2}\right\} -\frac{Mc}{\hbar}\right]\psi=0,\label{eq:25}
\end{equation}
with the wave function $\psi$ is chosen as
\begin{equation}
\psi=e^{-\frac{iEt}{\hbar}}\left(\psi_{1},\psi_{2},\psi_{3},\psi_{4},\psi_{5}\right)^{T}.\label{eq:26}
\end{equation}
Inserting (\ref{eq:26}) into Eq. (\ref{eq:25}), and after an algebraic
calculation, we obtain the following system of equations
\begin{equation}
\frac{1}{c\hbar}\left(E-\frac{kq}{r}\right)\psi_{2}-i\left(\partial_{r}+\frac{2}{r}\right)\psi_{3}-\frac{i}{r}\left(\partial_{\theta}+\frac{\text{cos}\theta}{\text{sin}\theta}\right)\psi_{4}-\frac{i}{a'r\text{sin}\theta}\left(\partial_{\varphi}-i\alpha\right)\psi_{5}=\frac{Mc}{\hbar}\psi_{1},\label{eq:27}
\end{equation}
\begin{equation}
\frac{1}{c\hbar}\left(E-\frac{kq}{r}\right)\psi_{1}=\frac{Mc}{\hbar}\psi_{2},\label{eq:28}
\end{equation}
\begin{equation}
i\partial_{r}\psi_{1}=\frac{Mc}{\hbar}\psi_{3},\label{eq:29}
\end{equation}
\begin{equation}
\frac{i}{r}\partial_{\theta}\psi_{1}=\frac{Mc}{\hbar}\psi_{4},\label{eq:30}
\end{equation}
\begin{equation}
\frac{i}{a'r\text{sin}\theta}\left(\partial_{\varphi}-i\alpha\right)\psi_{1}=\frac{Mc}{\hbar}\psi_{5}.\label{eq:31}
\end{equation}
Putting Eqs. (\ref{eq:28}), (\ref{eq:29}), (\ref{eq:30}) and (\ref{eq:31})
into Eq. (\ref{eq:27}), we have
\begin{equation}
\left\{ \frac{d^{2}}{dr^{2}}+\frac{2}{r}\frac{d}{dr}+\frac{1}{\left(c\hbar\right)^{2}}\left(E-\frac{kq}{r}\right)^{2}-\left(\frac{Mc}{\hbar}\right)^{2}-\frac{L^{2}}{r^{2}}\right\} \psi_{1}=0,\label{eq:32}
\end{equation}
where
\begin{equation}
L^{2}=-\hbar^{2}\left\{ \frac{1}{\text{sin}\theta}\frac{d}{d\theta}\left(\text{sin}\theta\frac{d}{d\theta}\right)+\frac{1}{a'^{2}\text{sin}^{2}\theta}\left(\partial_{\varphi}-i\alpha\right)^{2}\right\} ,\label{eq:33}
\end{equation}
\begin{equation}
L_{z}=-i\hbar\left(\frac{d}{d\varphi}-i\alpha\right),\label{eq:34}
\end{equation}
and where
\begin{equation}
l=\left|\lambda\right|+\kappa,\,\left(\kappa=0,1,2,\ldots\right),\label{eq:35}
\end{equation}
\begin{equation}
\lambda=\frac{m-\alpha}{a'},\label{eq:36}
\end{equation}
with $l$ are the eigenvalues of $L_{z}$ and $L^{2}$ respectively:
we note\textcolor{red}{{} that (i) these eigenvalues depend on the amount
$\alpha$ of the magnetic flux and the geometric parameter of space
$a'$, and (ii) they are not integer whatever the nature of the parameter
$\alpha$. }

In order to solve Eq. (\ref{eq:32}), we first put that
\begin{equation}
\psi_{1}\left(r,\theta,\phi\right)=R\left(r\right)\chi\left(\theta,\varphi\right).\label{eq:37}
\end{equation}
Expanding Eq. (\ref{eq:32}) leads to
\begin{equation}
\left\{ \frac{d^{2}}{dr^{2}}+\frac{2}{r}\frac{d}{dr}-\frac{l\left(l+1\right)-\left(\frac{kq}{\hbar c}\right)^{2}}{r^{2}}+\frac{E^{2}-M^{2}c^{4}}{\hbar^{2}c^{2}}-\frac{2kqE}{\hbar^{2}c^{2}}\frac{1}{r}\right\} R\left(r\right)=0.\label{eq:38}
\end{equation}
By using the following substitutions
\begin{equation}
\rho=\xi r,\,\xi^{2}=\frac{4\left(M^{2}c^{4}-E^{2}\right)}{\hbar^{2}c^{2}},\label{eq:39}
\end{equation}
\begin{equation}
\gamma=\frac{kq}{\hbar c},\,\varsigma=\frac{2\gamma E}{\hbar c\xi},\label{eq:40}
\end{equation}
Eq. (\ref{eq:38}) becomes
\begin{equation}
\left\{ \frac{1}{\rho^{2}}\frac{d}{d\rho}\left(\rho^{2}\frac{d}{d\rho}\right)-\frac{l\left(l+1\right)-\gamma^{2}}{\rho^{2}}-\frac{1}{4}-\frac{\varsigma}{\rho}\right\} R\left(\rho\right)=0.\label{eq:41}
\end{equation}
Putting that
\begin{equation}
R\left(\rho\right)=e^{-\frac{\rho}{2}}\rho^{s}u\left(\rho\right),\label{eq:42}
\end{equation}
Eq. (\ref{eq:41}) is transformed into
\begin{equation}
\rho\frac{d^{2}u\left(\rho\right)}{d\rho^{2}}+\left\{ 2\left(s+1\right)-\rho\right\} \frac{du\left(\rho\right)}{d\rho}-\left(s+\varsigma+1\right)u\left(\rho\right)=0,\label{eq:43}
\end{equation}
with 
\begin{equation}
s\left(s+1\right)=l\left(l+1\right)-\gamma^{2},\label{eq:44}
\end{equation}
or
\begin{equation}
s=-\frac{1}{2}+\sqrt{\left(l+\frac{1}{2}\right)^{2}-\gamma^{2}}.\label{eq:45}
\end{equation}
In this stage the solutions are
\begin{equation}
u\left(\rho\right)=N_{\text{norm}}\,_{1}F_{1}\left(s+\varsigma+1;2\left(s+1\right),\rho\right),\label{eq:46}
\end{equation}
with $N_{\text{norm}}$ is the normalization constant and $_{1}F_{1}$
is the confluent series.

The confluent series becomes a polynomial if and only if \citep{23}
\begin{equation}
s+\varsigma+1=-n,\,\left(n=0,1,2,\ldots\right).\label{eq:47}
\end{equation}
Consequently, the energy of spectrum is written as
\begin{equation}
E_{n}=\frac{Mc^{2}}{\sqrt{1+\frac{\gamma^{2}}{\left(n+\frac{1}{2}+\sqrt{\left(\left|\frac{m-\alpha}{a'}\right|+\kappa+\frac{1}{2}\right)^{2}-\gamma^{2}}\right)^{2}}}}.\label{eq:48}
\end{equation}
Now, concerning the total wave function, and according to the work
of \citep{8}, we have two cases 
\begin{itemize}
\item in the subspace $S_{+}$, where $\lambda_{1}>0$ and $l_{1}=\lambda_{1}+\kappa$
, the total spinor is
\begin{equation}
\psi_{n\lambda_{1}l_{1}}=\frac{\hbar}{Mc}e^{-\frac{iEt}{\hbar}}e^{im\varphi}\left(\begin{array}{c}
\frac{Mc}{\hbar}\\
\frac{1}{c\hbar}\left(E-\frac{kq}{r}\right)\\
i\partial_{r}\\
\frac{i}{r}\partial_{\theta}\\
\frac{i}{a'r\text{sin}\theta}\left(\partial_{\varphi}-i\alpha\right)
\end{array}\right)\psi_{1},
\end{equation}
with
\begin{equation}
\psi_{1}=N_{\text{norm}}P_{l_{1}}^{-\lambda_{1}}\left(\text{cos}\theta\right)\rho^{s}e^{-\frac{\rho}{2}}\,_{1}F_{1}\left(s+\varsigma+1;2\left(s+1\right),\rho\right).
\end{equation}
\item for the subspace $S_{-}$, where $\lambda_{2}<0$ and $l_{2}=-\lambda_{2}+\kappa$
, the total spinor is
\begin{equation}
\psi_{n\lambda_{2}l_{2}}=\frac{\hbar}{Mc}e^{-\frac{iEt}{\hbar}}e^{im\varphi}\left(\begin{array}{c}
\frac{Mc}{\hbar}\\
\frac{1}{c\hbar}\left(E-\frac{kq}{r}\right)\\
i\partial_{r}\\
\frac{i}{r}\partial_{\theta}\\
\frac{i}{a'r\text{sin}\theta}\left(\partial_{\varphi}-i\alpha\right)
\end{array}\right)\psi_{1},
\end{equation}
with
\begin{equation}
\psi_{1}=N_{\text{norm}}P_{l_{2}}^{\lambda_{2}}\left(\text{cos}\theta\right)\rho^{s}e^{-\frac{\rho}{2}}\,_{1}F_{1}\left(s+\varsigma+1;2\left(s+1\right),\rho\right).
\end{equation}
\end{itemize}

\section{The dkp equation with the aharonov-bohm and coulomb potentials in
the presence of a global monopole.}

\subsection{The solutions}

The metric of the space-time in the presence of a global monopole
is given by \citep{21,22,37,39,40}
\begin{equation}
ds^{2}=dt^{2}-dr^{2}-b'^{2}r^{2}\left(d\theta^{2}+\text{sin}^{2}\theta d\varphi^{2}\right).\label{eq:49}
\end{equation}
By using Eqs. (\ref{eq:3}), (\ref{eq:6}) and (\ref{eq:13}), the
DKP equation with the \textbf{AB} and Coulomb potentials is written
as
\begin{equation}
\left[\frac{\beta^{0}}{\hbar c}\left(E-\frac{kq}{r}\right)+i\beta^{1}\partial_{r}+\frac{i\beta^{2}}{br}\left(\partial_{\theta}-b'\beta^{2}\beta^{1}\right)+\frac{i\beta^{3}}{b'r\text{sin}\theta}\left\{ \left(\partial_{\varphi}-i\alpha\right)-b'\text{sin}\theta\beta^{3}\beta^{1}-\text{cos}\theta\beta^{3}\beta^{2}\right\} -\frac{Mc}{\hbar}\right]\varPsi=0.\label{eq:52}
\end{equation}
with\textcolor{red}{
\[
\varPsi=e^{-i\frac{Et}{\hbar}}\left(\varPsi_{1},\varPsi_{2},\varPsi_{3},\varPsi_{4},\varPsi_{5}\right)^{T},
\]
}and with the 4-vector potential, in the background of a global monopole,
reads\textcolor{red}{
\begin{equation}
A_{0}=\frac{kq}{r},\,A_{r}=0,A_{\theta}=0,\,A_{\varphi}=\frac{\Phi}{2\pi b'r\text{sin}\theta}.\label{eq:52.1}
\end{equation}
}Putting (\ref{eq:3}) in (\ref{eq:52}), we obtain the following
system of equations
\begin{equation}
\frac{1}{c\hbar}\left(E-\frac{kq}{r}\right)\varPsi_{2}-i\left(\partial_{r}+\frac{2}{r}\right)\varPsi_{3}-\frac{i}{br}\left(\partial_{\theta}+\frac{\text{cos}\theta}{\text{sin}\theta}\right)\varPsi_{4}-\frac{i}{b'r\text{sin}\theta}\left(\partial_{\varphi}-i\alpha\right)\varPsi_{5}=\frac{Mc}{\hbar}\varPsi_{1},\label{eq:53}
\end{equation}
\begin{equation}
\frac{1}{c\hbar}\left(E-\frac{kq}{r}\right)\varPsi_{1}=\frac{Mc}{\hbar}\varPsi_{2},\label{eq:54}
\end{equation}
\begin{equation}
i\partial_{r}\varPsi_{1}=\frac{Mc}{\hbar}\varPsi_{3},\label{eq:55}
\end{equation}
\begin{equation}
\frac{i}{br}\partial_{\theta}\varPsi_{1}=\frac{Mc}{\hbar}\varPsi_{4},\label{eq:56}
\end{equation}
\begin{equation}
\frac{i}{b'r\text{sin}\theta}\left(\partial_{\varphi}-i\alpha\right)\varPsi_{1}=\frac{Mc}{\hbar}\varPsi_{5}.\label{eq:57}
\end{equation}
Now, inserting Eqs. (\ref{eq:54}), (\ref{eq:55}), (\ref{eq:56})
and (\ref{eq:57}) into Eq. (\ref{eq:53}), we obtain
\begin{equation}
\left\{ \frac{d^{2}}{dr^{2}}+\frac{2}{r}\frac{d}{dr}+\frac{1}{\left(c\hbar\right)^{2}}\left(E-\frac{kq}{r}\right)^{2}-\left(\frac{Mc}{\hbar}\right)^{2}-\frac{L^{2}}{r^{2}}\right\} \varPsi_{1}=0,\label{eq:57-1}
\end{equation}
where
\begin{equation}
L^{2}=-\hbar^{2}\left\{ \frac{1}{\text{sin}\theta}\frac{d}{d\theta}\left(\text{sin}\theta\frac{d}{d\theta}\right)+\frac{1}{b'^{2}\text{sin}^{2}\theta}\left(\partial_{\varphi}-i\alpha\right)^{2}\right\} ,\label{eq:57-2}
\end{equation}
\begin{equation}
L_{z}=-i\hbar\left(\frac{d}{d\varphi}-i\alpha\right),\label{eq:57-3}
\end{equation}
and
\begin{equation}
l=\left|\lambda\right|+\kappa,\,\left(\kappa=0,1,2,\ldots\right)\label{eq:57-4}
\end{equation}
with
\begin{equation}
\lambda=\frac{m-\alpha}{b'},\label{eq:57-5}
\end{equation}
and $l$ are the eigenvalues of $L_{z}$ and $L^{2}$ respectively:\textcolor{red}{{}
we observe that (i) these eigenvalues depend on the amount $\alpha$
of the magnetic flux and the geometric parameter of space $b'$, and
(ii) they are not integer whatever the nature of the parameter $\alpha$. }

In order to solve Eq. (\ref{eq:57-1}), we first put that
\begin{equation}
\varPsi_{1}\left(r,\theta,\phi\right)=R\left(r\right)\chi\left(\theta,\varphi\right).\label{eq:37-1}
\end{equation}
Expanding Eq. (\ref{eq:57-1}) leads to

\begin{equation}
\left\{ \frac{d^{2}}{dr^{2}}+\frac{2}{r}\frac{d}{dr}+\frac{1}{\left(c\hbar\right)^{2}}\left(E-\frac{kq}{r}\right)^{2}-\left(\frac{Mc}{\hbar}\right)^{2}-\frac{\frac{l\left(l+1\right)}{b'^{2}}}{r^{2}}\right\} \varPsi_{1}\left(r\right)=0,\label{eq:58}
\end{equation}
Using the same change of variables as in the case of cosmic string,
Eq. (\ref{eq:58}) is transformed into
\begin{equation}
\frac{d^{2}R\left(r\right)}{d\rho^{2}}+\left(-\frac{\frac{l\left(l+1\right)}{b'^{2}}-\gamma^{2}}{\rho^{2}}-\frac{\zeta}{\rho}-\frac{1}{4}\right)R\left(r\right)=0.\label{eq:60}
\end{equation}
Now let us make a change of variable
\begin{equation}
R\left(r\right)=\rho^{a}e^{-\frac{\rho}{2}}H\left(\rho\right),\label{eq:61}
\end{equation}
and putting it in (\ref{eq:60}), we have
\[
\rho^{2}\frac{d^{2}H\left(\rho\right)}{d\rho^{2}}+\left\{ 2\left(a+1\right)\rho-\rho^{2}\right\} \frac{dH\left(\rho\right)}{d\rho}
\]
\begin{equation}
+\left[\left\{ a\left(a+1\right)-\left(\frac{l\left(l+1\right)}{b'^{2}}-\gamma^{2}\right)\right\} -\left(\zeta+a+1\right)\rho\right]H\left(\rho\right)=0,\label{eq:62}
\end{equation}
with
\begin{equation}
a=-\frac{1}{2}\pm\left\{ \frac{l\left(l+1\right)}{b'^{2}}+\frac{1}{4}-\gamma^{2}\right\} .^{\frac{1}{2}}\label{eq:63}
\end{equation}
To solve Eq. (\ref{eq:62}), we use the Frobenius method \citep{24,25,26,27},
This can be written as a power series expansion around the origin:
\begin{equation}
H\left(\rho\right)=\sum_{k=0}^{\infty}c_{k}\rho^{k},\label{eq:64}
\end{equation}
Inserting (\ref{eq:64}) into (\ref{eq:62}), we obtain the following
recurrence relation:
\begin{equation}
c_{k+1}=\frac{k+\left(\zeta+a+1\right)}{k\left(k+1\right)+2\left(a+1\right)\left(k+1\right)}c_{k}.
\end{equation}
Now, when $k\rightarrow\infty$, the ration $\frac{c_{k+1}}{c_{k}}\rightarrow0$:
we can understand this condition by saying that: Special kind of exact
solutions, which represent bound states, can be obtained looking for
polynomials expressions to $H(\rho)$, i.e., the solutions can be
obtained by imposing the conditions where power series becomes a polynomial
of degree $n$. We can be argued this as follows: in quantum mechanics,
if we want to have a normalizable wave function, we have to impose
that $R(\rho)$ vanishes at $\rho\rightarrow0$ and $\rho\rightarrow\infty$
. In this way, bound state solutions can be obtained because there
is no divergence of the wave function at $\rho\rightarrow0$ and $\rho\rightarrow\infty$
. In our case, the $H(\rho)$ has written as a power series expansion
around the origin (see Eq. (\ref{eq:64})). As a result of that, the
solutions can be achieved by imposing that the power series expansion
(\ref{eq:64}) or the biconfluent Heun series becomes a polynomial
of degree$n$ . This guaranteeing that $R(\rho)$ behaves $F$ as
$\rho$ at the origin and vanishes at $\rho\rightarrow\infty$ \citep{28,29,30,31,32,33}.
Thus, in order that the power series expansion becomes a polynomial
of degree $n$, we impose that
\begin{equation}
\zeta+a+1=-n.
\end{equation}
In this case, the form of spectrum of energy is 
\begin{equation}
E_{n}=Mc^{2}\left\{ 1+\frac{\gamma^{2}}{\left[n+\frac{1}{2}+\left\{ \frac{l\left(l+1\right)}{b'^{2}}+\frac{1}{4}-\gamma^{2}\right\} ^{\frac{1}{2}}\right]^{2}}\right\} ^{-\frac{1}{2}}.
\end{equation}
Now, concerning the total wave function, we have also two cases \citep{8}
\begin{itemize}
\item in the subspace $S_{+}$, where $\lambda_{1}>0$ and $l_{1}=\lambda_{1}+\kappa$
, the total spinor is
\begin{equation}
\varPsi_{n\lambda_{1}l_{1}}=\frac{\hbar}{Mc}e^{-\frac{iEt}{\hbar}}e^{im\varphi}\left(\begin{array}{c}
\frac{Mc}{\hbar}\\
\frac{1}{c\hbar}\left(E-\frac{kq}{r}\right)\\
i\partial_{r}\\
\frac{i}{r}\partial_{\theta}\\
\frac{i}{b'r\text{sin}\theta}\left(\partial_{\varphi}-im_{0}\right)
\end{array}\right)\psi_{1},
\end{equation}
with
\begin{equation}
\psi_{1}=N'_{\text{norm}}P_{l_{1}}^{-\lambda_{1}}\left(\text{cos}\theta\right)\rho^{a}e^{-\frac{\rho}{2}}\,_{1}F_{1}\left(a+\varsigma+1;2\left(a+1\right),\rho\right).
\end{equation}
\item for the subspace $S_{-}$, where $\lambda_{2}<0$ and $l_{2}=-\lambda_{2}+\kappa$
, the total spinor is
\begin{equation}
\varPsi_{n\lambda_{2}l_{2}}=\frac{\hbar}{Mc}e^{-\frac{iEt}{\hbar}}e^{im\varphi}\left(\begin{array}{c}
\frac{Mc}{\hbar}\\
\frac{1}{c\hbar}\left(E-\frac{kq}{r}\right)\\
i\partial_{r}\\
\frac{i}{r}\partial_{\theta}\\
\frac{i}{b'r\text{sin}\theta}\left(\partial_{\varphi}-i\alpha\right)
\end{array}\right)\psi_{1},
\end{equation}
with
\begin{equation}
\psi_{1}=N'_{\text{norm}}P_{l_{2}}^{\lambda_{2}}\left(\text{cos}\theta\right)\rho^{a}e^{-\frac{\rho}{2}}\,_{1}F_{1}\left(a+\varsigma+1;2\left(a+1\right),\rho\right).
\end{equation}
\end{itemize}
Here, $N'_{\text{norm}}$ is the normalization constant and $_{1}F_{1}$
is the confluent series.

\section{Conclusion}

\textcolor{red}{This paper is devoted to study the solutions of the
relativistic quantum motion of a charged scalar particles in the presence
of a Aharonov\textendash Bohm and Coulomb potentials in the spacetimes
produced by an idealized cosmic string and global monopole. These
solutions have been obtained, and the influence of the parameter of
the geometry of both spaces has been discussed. In addition, the remarks
which Chun-Fang Li \citep{11,12} has been proposed concerning the
}\textbf{\textcolor{red}{AB}}\textcolor{red}{{} effect, have been extended
in our case: thus, the presence of }\textbf{\textcolor{red}{AB}}\textcolor{red}{{}
potential changes completely the fundamental commutation relations
of the angular momentum. Following the works of \citep{11,12}, we
note that (i) the KAM are not satisfied even when the particle is
restricted to the doubly-connected space where it does not touch the
magnetic field on the z-axis, (ii) the region where the magnetic field
exists and is inaccessible to the electron should be taken into account
in the physically commutation relations, and finally (iii) the Pauli
criterion which said that \textquotedblleft the appropriate eigenfunctions
are those which are square integrable and are closed under the operation
of ladder operators \textquotedblleft{} in inapplicable to the vector
}\textbf{\textcolor{red}{AB}}\textcolor{red}{. The existence of the
magnetic field on the $z-$axis is the principal cause of breaking
down the symmetry of particle's motion around the $z-$axis.}

\textcolor{red}{The eigenfunctions and eigenvalues of $L_{z}$ and
$L^{2}$ have been presented under the following boundary condition
$\left.\psi\left(r,\theta,\varphi\right)\right|_{\theta=0,\pi}=0$,
and the space $S$, as showed by \citep{8}, is split into two subspaces,
$S_{+}$ and $S_{-}$: $S_{+}$ is spanned by all the wave function
$\psi_{nl_{1}\lambda_{1}}\left(\theta,\varphi\right)$ for cosmic
string ($\varPsi_{nl_{1}\lambda_{1}}\left(\theta,\varphi\right)$
for global monopole) and $S_{-}$ is spanned by all the wave function
$\psi_{nl_{2}\lambda_{2}}\left(\theta,\varphi\right)$ for cosmic
string ($\varPsi_{nl_{2}\lambda_{2}}\left(\theta,\varphi\right)$
for global monopole). The eigensolutions in the gravitational field
of cosmic string and a global monopole (i) are different, (ii) depend
on the geometry of spaces.}

\end{document}